\documentclass[showpacs,superscriptaddress,twocolumn,aps]{revtex4}
\usepackage{amssymb}
\usepackage{graphicx}
\usepackage{dcolumn}
\usepackage{bm}
\usepackage{amsmath,amssymb}
\usepackage{graphicx}
\usepackage{color}

\begin{document}

\title{The fidelity of general bosonic channels with pure state input}
\author{Meisheng Zhao}
\email{zmesson@mail.ustc.edu.cn}
\affiliation{Department of Modern Physics, University of Science and Technology of China,
Hefei 230026, People's Republic of China}
\author{Tao Qin}
\affiliation{Department of Modern Physics, University of Science and Technology of China,
Hefei 230026, People's Republic of China}
\author{Yongde Zhang}
\affiliation{CCAST (World Laboratory), P.O. Box 8730, Beijing 100080, People's Republic
of China}
\affiliation{Department of Modern Physics, University of Science and Technology of China,
Hefei 230026, People's Republic of China}
\date{\today}

\begin{abstract}
We first derive for the general form of the fidelity for various bosonic
channels. Thereby we give the fidelity of different quantum bosonic channel,
possibly with product input and entangled input respectively, as examples.
The properties of the fidelity are carefully examined.
\end{abstract}

\maketitle

Quantum bosonic channels are a specific type of quantum channels with
continuous alphabet \cite{rmp}. Due to their important applications to
optical transmission in which photons are employed to convey the
information, large efforts have been devoted to studying the properties of
these kinds of channels. For example, the calculation for the quantum and
classical capacities, entropy, or fidelity has drawn much attention.
Particularly, the fidelity can be applied to measure how close the input
states and the output states of the quantum channels are. A fidelity of
unity implies identical quantum states while a fidelity of zero implies
orthogonal quantum states. To some extents, it evaluates how well the
quantum channels preserve the transmitted information thus it is an
essential physical quantity in quantum information theory \cite{wang,caves}.

The evaluation of the fidelity for a class of bosonic channels might be also
relevant for quantum tasks such as continuous variables cloning and
teleportation \cite{caves}.

Specifically, here we study the fidelity of general quantum bosonic Gaussian
channels on occasions input is a pure Gaussian state. These bosonic channels
represent multiple scenarios of quantum information transmission, thus
render our study into a nontrivial one.

A quantum system with $n$ modes described by $n$ pairs of canonical
coordinates $(x_{i},p_{i})$ is a CCR(canonical commute relation) system\cite%
{Lindblad,Demoen}, and the annihilation and creation operators $(a_{i},\
a_{i}^{+})$ of these modes are related to $(x_{i},p_{i})$ according to $%
x_{i}=\frac{1}{\sqrt{2}}(a_{i}+a_{i}^{+}),\ p_{i}=\frac{-i}{\sqrt{2}}%
(a_{i}-a_{i}^{+})$. For such a system, a state $\rho $ characterized by its
Weyl-Wigner distribution,
\begin{equation*}
\varkappa (\rho _{\varepsilon })=tr\left( \rho W_{\varepsilon }\right)
\end{equation*}%
where $W_{\varepsilon }=exp\{i\varepsilon R^{T}\}$ is the Weyl operator.
Here $\varepsilon $ is 2n dimensional real vector, $%
R=(x_{1},x_{2},...,x_{n},p_{1},p_{2},...,p_{n})$. the state can be obtained
as $\rho =\left( 2\pi \right) ^{-n}\int \varkappa (\rho _{\varepsilon
})W\left( \varepsilon \right) ^{+}d^{2n}\varepsilon $. A Gaussian state have
a Gaussian Weyl-Wigner distribution and thus can be written as below:
\begin{equation*}
\rho _{G}=\left( 2\pi \right) ^{-n}\int e^{-\frac{1}{2}\varepsilon \Gamma
\varepsilon ^{T}+iD\varepsilon ^{T}}W\left( \varepsilon \right)
^{+}d^{2n}\varepsilon
\end{equation*}%
in which the $2n\times 2n$ matrix $\Gamma $ is the covariance matrix and $2n$
dimensional real vector $D$ are the displacements with defination $%
D_{i}=tr(R_{i}\rho ),\Gamma _{i,j}=Re\left\langle
(R_{i}-D_{i})(R_{j}-D_{j})\right\rangle _{\rho },i,j=(1,2,...,n)$. The
positivity of quantum states requires $\Gamma +i\sigma >=0,\ \sigma =\oplus
i\sigma _{2},$ and $\sigma _{2}$ is the second Pauli matrix, So $\Gamma >=0$
is also required. When the state is pure, $\Gamma >0\subseteq Sp(2n)$.

A general Gaussian channel $\rho \longrightarrow T(\rho)$ can be defined by
its action on weyl operator \cite{wolf,holevo},%
\begin{eqnarray*}
T(W_{\varepsilon})\longrightarrow W_{\varepsilon A }^{+} e^{-\frac{1}{2}%
\varepsilon G \varepsilon^{T}}
\end{eqnarray*}%
where $A,G$ is $2n\times2n$ real matrices, and $G>0$ should be symmetric
matrix. Then we obtain output state
\begin{eqnarray*}
\rho^{out}=\frac{1}{\left( 2\pi \right) ^{n}}\int e^{-\frac{1}{2}\varepsilon
(A \Gamma A^{T}+G)\varepsilon ^{T}+iDA^{T}\varepsilon^{T} }W\left(
\varepsilon \right) d^{2n}\varepsilon
\end{eqnarray*}

Therefore, the input-output fidelity of a general Gaussian channel with
Gaussian pure input is
\begin{widetext}
\begin{eqnarray*}
\digamma  &=&Tr\left( \rho _{in}\rho _{out}\right)  \\
&=&(2\pi )^{-n}\int e^{-\frac{1}{2}\varepsilon (A\Gamma A^{T}+G)\varepsilon
^{T}+iDA^{T}\varepsilon ^{T}}Tr\left( \rho _{in}w^{+}\left( \varepsilon
\right) \right) d^{2n}\varepsilon  \\
&=&(2\pi )^{-n}\int e^{-\frac{1}{2}\varepsilon (A\Gamma A^{T}+\Gamma
+G)\varepsilon ^{T}+iD(A^{T}-I)\varepsilon ^{T}}d^{2n}\varepsilon  \\
&=&\frac{1}{\det \sqrt{A\Gamma A^{T}+\Gamma +G}}\exp \left\{ -\left(
DA^{T}-D\right) \frac{1}{2A\Gamma A^{T}+2\Gamma +2G}\left( DA^{T}-D\right)
^{T}\right\}
\end{eqnarray*}
\end{widetext}
As $\Gamma>0$ and $G>0$, so the displacements $D$ will not increase the
fidelity.

We will present some examples in next steps.

The first important example of Gaussian channel is the single-mode
amplification channel, for which $A=\sqrt{\eta }\mathbf{\ I}_{2},\ G=(\eta
-1)\mathbf{\ I}_{2}$, $\mathbf{I}_{2}$ is the identity matrix of two
dimension. Amplification channel has important application in C-V cloning
process. For this kind of channel,we can give the fidelity directly
\begin{equation*}
\digamma =(det\sqrt{(1+\eta )\Gamma +\eta -1})^{-1}
\end{equation*}%
the maximum $\digamma =2/(3\eta -1)$ reaches at $\Gamma =\mathbf{I}/2$.

Another important example is the classical noise channel for which $A=I,G>0$%
. For these channels displacements have no effect on fidelity, which is a
good property for process like cloning or teleportation. The Gaussian C-V
clone and a large class of C-V teleportation both can just be described as a
single mode classical noise channel.

We will consider the occasion a bit more complex when there exists
correlated noise.

Now memory arises in bosonic Gaussian channel. The bosonic Gaussian memory
channel is characterized by the following map \cite{macchi}%
\begin{widetext}
\begin{equation*}
\$:\$\left( \rho _{in}\right) =\int d^{2}\beta _{1}d^{2}\beta _{2}q\left(
\beta _{1},\beta _{2}\right) D\left( \beta _{1}\right) \otimes D\left( \beta
_{2}\right) \rho _{in}D^{+}\left( \beta _{1}\right) \otimes D^{+}\left(
\beta _{2}\right)
\end{equation*}%
\end{widetext}
with%
\begin{equation*}
q\left( \beta _{1},\beta _{2}\right) =\frac{1}{\pi ^{2}\sqrt{\left\vert
\gamma _{N}\right\vert }}e^{-\beta ^{+}\gamma _{N}^{-1}\beta }
\end{equation*}%
where $\beta =\left[ \Re \left( \beta _{1}\right) ,\Im \left( \beta
_{1}\right) ,\Re \left( \beta _{2}\right) ,\Im \left( \beta _{2}\right) %
\right] ^{T}$ and $\gamma _{N}$ is the covariance matrix of the noise
quadratures%
\begin{equation*}
\gamma _{N}=\left(
\begin{array}{cccc}
N & 0 & -xN & 0 \\
0 & N & 0 & xN \\
-xN & 0 & N & 0 \\
0 & xN & 0 & N%
\end{array}%
\right)
\end{equation*}%
where $x$ is the correlation coefficient ranging from $0$ to $1$. When $x=1$%
, the channel is with full memory; when $x=0$, the channel is memoryless.we
can easily rewrite this channel in Heisenberg picture:
\begin{equation*}
\$\left( W_{\varepsilon} \right)=W_{\varepsilon} e^{\varepsilon \gamma_{N}
\varepsilon^{T}}
\end{equation*}

Assume the input state is product coherent state%
\begin{equation*}
\rho _{in}=\left\vert \gamma _{1}\gamma _{2}\right\rangle \left\langle
\gamma _{1}\gamma _{2}\right\vert
\end{equation*}

Then we have the following relations:%
\begin{equation*}
\Gamma =\frac{1}{2}
\end{equation*}

Finally the fidelity comes to be%
\begin{eqnarray}
\digamma _{Mem} &=&\frac{1}{\det \sqrt{2\Gamma +\gamma_{N}}}  \notag \\
&=&\frac{1}{\left( N+1\right) ^{2}-N^{2}x^{2}}
\end{eqnarray}

Remembering that $x$ ranges from $0$ to $1$, therefore the fidelity has
upper bound $\digamma _{Mem}^{Upp}$%
\begin{equation*}
\digamma _{Mem}^{Upp}=\frac{1}{\left( N+1\right) ^{2}}
\end{equation*}%
and lower bound $\digamma _{Mem}^{Low}$%
\begin{equation*}
\digamma _{Mem}^{Low}=\frac{1}{\left( N+1\right) ^{2}-N^{2}}
\end{equation*}

From the equations above, we can see that when the memory intensity $x$ of
this channel becomes larger, the fidelity also becomes larger. Besides, the
fidelity decreases with growing noise variance $N$.

Quantum entanglement has proven to be a valuable resource which has wide
applications in quantum information domain \cite{entanglement}. Its
intrinsic nonlocal nature often enables it to have better performance to
accomplish the quantum information tasks.

When the input state to the bosonic memory channel is entangled continuous
variable state, more complex results are expected.

Assume the input state is bipartite entangled vacuum state%
\begin{equation*}
\rho _{in}=S\left( \gamma \right) \left\vert 00\right\rangle =\frac{1}{\cosh
\gamma }e^{\tanh \gamma a_{1}^{+}a_{2}^{+}}\left\vert 00\right\rangle
\end{equation*}

In this case, we have $\Gamma $%
\begin{equation*}
\Gamma =\left(
\begin{array}{cccc}
\cosh 2r & 0 & -\sinh 2r & 0 \\
0 & \cosh 2r & 0 & \sinh 2r \\
-\sinh 2r & 0 & \cosh 2r & 0 \\
0 & \sinh 2r & 0 & \cosh 2r%
\end{array}%
\right)
\end{equation*}%
and
\begin{widetext}
\begin{equation*}
G=\gamma _{N}=\left(
\begin{array}{cccc}
\cosh 2r+N & 0 & -\sinh 2r-xN & 0 \\
0 & \cosh 2r+N & 0 & \sinh 2r+xN \\
-\sinh 2r-xN & 0 & \cosh 2r+N & 0 \\
0 & \sinh 2r+xN & 0 & \cosh 2r+N%
\end{array}%
\right)
\end{equation*}
\end{widetext}

Consequently, the fidelity is%
\begin{equation}
\digamma =\frac{1}{1+N^{2}+2N\cosh 2r-x^{2}N^{2}-2xN\sinh 2r}
\end{equation}

Examining Eq. (2), we find that same principles hold: the more noisy the
channel, the less the fidelity of the channel. Besides, the fidelity of the
channel increases with larger squeeze parameter $r$. And still, the fidelity
of the channel grows larger with increasing memory intensity $x$.

In this paper, we study partially the fidelity in cases\ of bosonic Gaussian
channel with a pure state input. In principle, the fidelity decreases with
noise variance $N$. When there is leakage of energy from the channel into
the environment, the fidelity drops off with passing time $t$.

With the presence of memory, the fidelity of the channel with product
coherent input state increases with growing memory intensity $x$. If the
input state to the bosonic memory channel is entangled squeezed vacuum
state, the fidelity of the channel also increases with growing squeeze
parameter $\gamma $, which represents how highly entangled the state is.

The studies here can be deepened by exploiting the fidelity of the channel
with more complex input states. Our study is limited to finite uses of the
bosonic channels. The situation for the fidelity of the channel of infinite
uses of the bosonic channels is expected to be more interesting.

The fidelity is a crucial physical quantity to evaluate the quality of
information transmission. We hope our research can make this issue more
clarified.


\begin{thebibliography}{9}
\bibitem{rmp} C. M. Caves and P. D. Drummond, Rev. Mod. Phys. \textbf{66},
481 (1994).

\bibitem{wang} Xiang-Bin Wang, L C Kwek and C H Oh, J. Phys. A: Math. Gen.
\textbf{33}, 4925 (2000).

\bibitem{caves} Carlton M. Caves and Krzysztof Wodkiewicz, \textit{Open Sys.
\& Information Dyn}. \textbf{11}, 309 (2006).

\bibitem{Demoen} B. Demoen, P. Vanheuswijn, and A. Verbeure, Lett. Math.
Phys. 2, 161 (1977).

\bibitem{Lindblad} G.Lindblad, J. Phys. A 33,59 (2000).

\bibitem{holevo} A.S. Holevo, Probabilistic Aspects of Quantum Theory
(North-Holland, Amsterdam, 1982), Chapter 5.

\bibitem{wolf} J.Eisert and M.M.Wolf, quant-ph/00505151.

\bibitem{macchi} N. J. Cerf, J. Clavareau, C. Macchiavello, and J. Roland,
Phys. Rev. A \textbf{72}, 042330 (2005).

\bibitem{entanglement} M. Nielsen and I. Chuang, \textit{Quantum Computation
and Quantum Information} (2000) Cambridge University Press, Cambridge; John
Preskill, http://www.theory.caltech.edu/\~{}preskill/ph229.
\end{thebibliography}
\end{document}